\documentclass[12pt]{article}
\usepackage{amsfonts}
\usepackage{graphicx}
\usepackage{amssymb}

\author{Tristan Cragnolini, Philippe Derreumaux, Samuela Pasquali
\footnote{Laboratoire de Biochimie Th\'eorique, UPR 9080 CNRS, samuela.pasquali@ibpc.fr}}
\title{Coarse-grained simulations of RNA and DNA duplexes}

\begin{document}

\maketitle

\abstract{Although RNAs play many cellular functions little is known about the dynamics and thermodynamics of these molecules.
In principle, all-atom molecular dynamics simulations can investigate these issues, but with current computer facilities, 
these simulations have been limited to small RNAs and to short times.

HiRe-RNA, a recently proposed high-resolution coarse-grained for RNA that captures many geometric details
such as base pairing and stacking, is able to fold RNA molecules to near-native structures in a short computational time.
So far it had been applied to simple hairpins, and here we present its application to duplexes of a couple dozen nucleotides 
and show how with our model and with Replica Exchange Molecular Dynamics (REMD) we can easily predict the correct double helix from 
a completely random configuration and study the dissociation curve. 
To show the versatility of our model, we present an application to a double stranded DNA molecule as well.

A reconstruction algorithm allows us to obtain full atom structures from the coarse-grained model. 
Through atomistic Molecular Dynamics (MD) we can compare the dynamics starting from a representative structure of 
a low temperature replica or from the experimental structure, and show how the two are statistically identical, highlighting
the validity of a coarse-grained approach for structured RNAs and DNAs.
}

\section{Introduction}

The study of non-coding RNA has been a very active area of research in recent years,
as new functions were reported.
Many of these RNAs have to fold into specific three-dimensional structures
to function in the cell.

Despite advances in structure determination, many questions remain open concerning
the dynamics and thermodynamics of these molecules.
Computational models can complement the experimental data by giving access
to the complete atomistic configurations by integration of Newton's equation of motion, using an appropriate
force field \cite{Sim2012}.
However, the application of such all-atom Molecular Dynamics (MD) simulations is
limited by the considerable computational requirements to integrate such trajectories long enough to sample many folding/unfolding events,
even for RNAs containing as little as a dozen nucleotides \cite{Bowman2008}.

In recent years, coarse-grained models have been proposed to alleviate this problem
by using simplified representations of biological molecules while striving
to keep the important details \cite{Hyeon2011}. 
Such models now exist for both RNA \cite{Xayaphoummine2005, Cao2005, Bernauer2011a, Sharma2008, Jonikas2009a, Morriss-Andrews2010} and DNA \cite{Ouldridge2011, Dans2010, Poulain2008, Hsu2012}.
The various models reflect the different questions for which they were developed, and no universal coarse-grained model exists.
The models differ in the number of beads representing the nucleotides, i.e. in the level of resolution, in the presence or absence of contraints that 
impose bounds or secondary structures, and even in the shape of the beads, that can go from spherical to elliptical.

HiRe-RNA is such a model for RNAs, whose resolution is high enough
to preserve many important geometric details, such as base pairing and stacking \cite{Pasquali2010}, 
without imposing preset pairings for the nucleotides. 
The resolution level, the absence of constraints, and the full flexibility, make HiRe-RNA suitable to study folding and unfolding events. 

The first application of HiRe-RNA concerned simple RNA hairpin topologies of a single molecule.
This paper presents an improved version of HiRe-RNA and applies it to double stranded RNAs as well as to
a double stranded DNA molecule.

For three sets of molecules, we first performed Replica-Exchange Molecular Dynamics (REMD) simulations starting from random configurations,
computed specific heat curves and extracted low-energy structures by clustering of the sampled conformations. 
The most populated clusters at low temperatures were then reconstructed into fully atomistic structures.

To test whether an atomistic simulation based on the structure we obtain via coarse-grained REMD behaves differently from a 
simulation starting from the experimental configuration, we performed regular atomistic MD using the AMBER99 forcefield on both 
reconstructed and experimental structures.

In section \ref{meth} we present the coarse-grained model, with an emphasis on the changes with respect 
to the original model presented in \cite{Pasquali2010}, as well as the reconstruction scheme that allows us to obtain back an atomistic structure.
In section \ref{sys} we present the different molecules we have studied, two RNA duplexes with non canonical 
pairings, and one DNA duplex. 
In section \ref{CGsim} we present the results of REMD for all systems and extract thermodynamic information such as the melting curve 
as well as the population distribution at various temperatures. 
In section \ref{FAsim} we compare the atomistic simulations for the reconstructed and the experimental molecules.

\section{Model} \label{meth}

\subsection*{Force field}
\begin{figure}
 \begin{center}
 \includegraphics[width=0.4\textwidth]{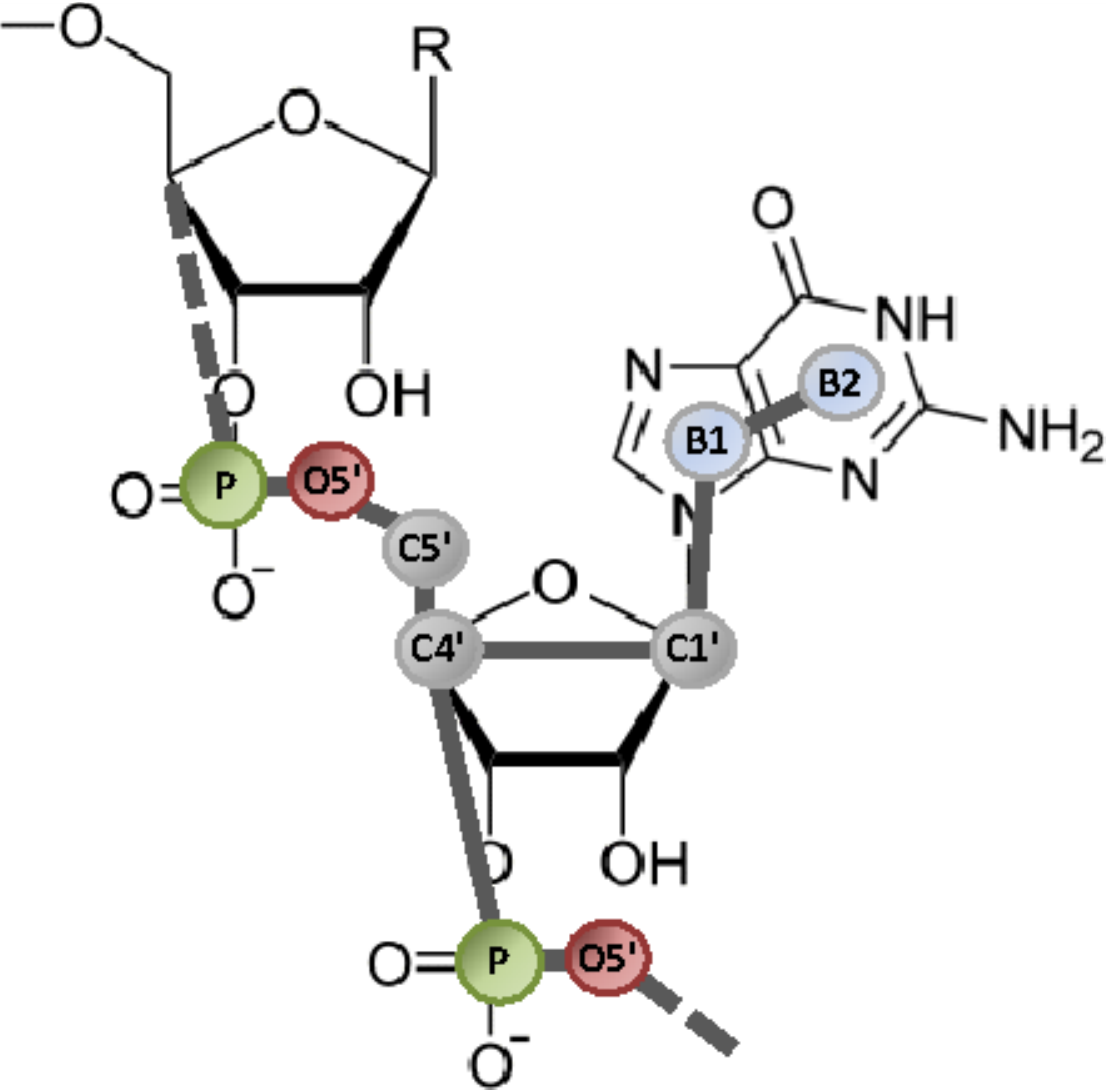}
 \caption{\label{fig_GG}HiRe-RNA coarse-grained representation for a guanine superposed on an all-atom representation.}
 \end{center}
\end{figure}

In our off-lattice HiRe-RNA model, each nucleotide is represented by six or seven beads (Fig. \ref{fig_GG}):
one bead for the phosphate - P -, four beads for the sugar - O5', C5', C4', C1'- one  
bead for the pyrimidine bases (B1 = C1 and U1) and two beads
for the purine bases (B1 = G1 or A1, B2 = G2 or A2). The positions of the B1 and B2 beads
coincide with the centers of mass of non-hydrogen atoms in the all-atom rings. 
The OH group specific to the ribose is not explicitly represented.
This feature, although limiting for RNA studies in some aspects, allows us to extend the model to DNA molecules as well
through a geometric reparametrization for equilibrium values of angles and torsions.

The HiRe-RNA force field is expressed as a sum of local, non-bonded and hydrogen-bond (H-bond) terms.
Local interactions are the usual harmonic terms for covalent bonds, angles, and torsions.
Non-bonded interactions are modelled by a modified Lennard-Jones, characterized by a power-law repulsion core at short distances,
a well width changing with the equilibrium distance, and an exponential tail at large distances.
Hydrogen bond interactions are specific of our model and consist of 3 terms: a 2-body interaction based on distance and angles between the
interacting beads, a 3-body term with the role of a repulsion barrier to avoid multiple H-bonds of just one base, and a 4-body term
of cooperativity involving successive base pairs. 
The 3-body and 4-body term enter in action only when the 2-body term reaches significant energies.

\subsubsection*{HiRe-RNA version $\beta$.2}
Leaving unaltered the local and non-boded terms presented in \cite{Pasquali2010},
we have introduced some modifications to the hydrogen-bond terms.
H-bonds are clearly the hardest interactions to account for because they are somewhere in between a classical and 
a quantum interaction. 
As we apply HiRe-RNA to more and more structures, we are inevitably brought to make changes to ameliorate this term.

\begin{figure}
  \begin{center}
  \includegraphics[width=0.6\textwidth]{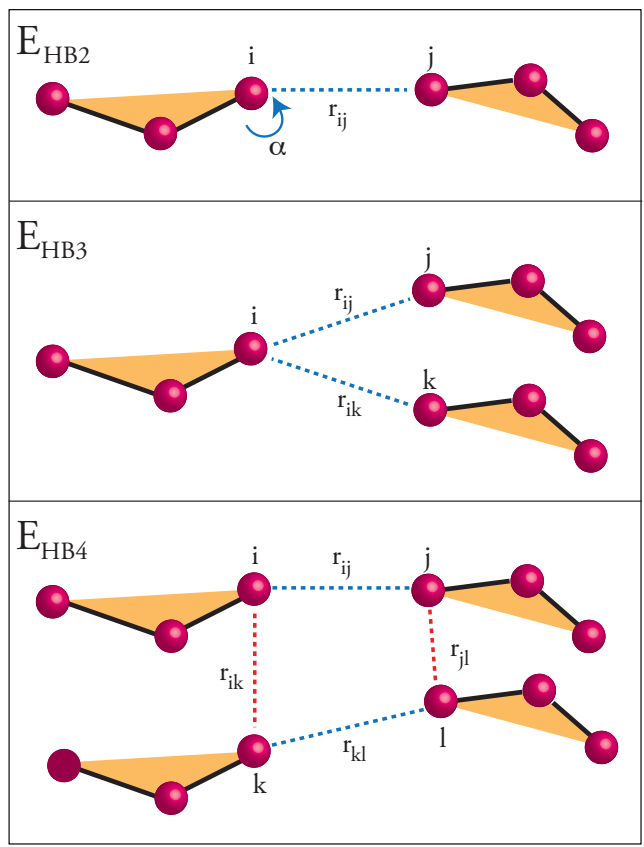}
  \caption{\label{fig_HB}Base pairing is modeled through three different energy terms that depend on the geometric configurations adopted by two ($E_{HB2}$), three ($E_{HB3}$), or four ($E_{HB4}$) bases. For each of these interactions we indicate the variables upon which the energy term depends.}
  \end{center}
\end{figure}

The two significant changes we introduced are in the 2-body and in the 4-body terms (refer to Figure \ref{fig_HB} for the general base-pairing scheme and 
particle numbering).
In the original version of the force field, the distance component of the 2-body term was a 5-10 LJ interaction. 
We have found that the repulsion core of the interaction was sometimes interfering with the stacking of successive base pairs.
Given the interaction grows spherically all around the interacting bead, as opposed to being confined on an hypothetical H-bond plane, the repulsion
could push away a bead interacting with the successive base.
The problem could rise for instance in the following scenario.
Consider a base $i$ interacting with a base $j$ and a base $i+1$ interacting with base a $j-1$. Depending on the overall torsion of the helix, it can occur
that the equilibrium distance between $i$ and $j+1$ is in a range where the repulsion core enters in action. Given that most likely the angular dependence of the
energy function is still rather good (since particle $j+1$ is aligned only a few \AA~ away from $j$ which bonds correctly with $i$), the dominant effect will be that
of repelling base $j+1$ away from $i$ and therefore stretching the helix. 
We decided to eliminate the repulsive core from the 2-body term of the hydrogen interaction and to make it exclusively attractive. 
A generic hard core repulsion between all beads in the non-bonded interaction already governs the excluded volume of all particles.

The distance contribution of the 2-body term takes now the simple form of an inverted Gaussian:
\begin{equation}
 E_{HB2} = -\varepsilon_{hb2}~ e^{-\nu (r_{i,j}^2 - \rho^2)^2}
\end{equation}
where $\varepsilon_{hb2}$ is the strength of the interaction, $\nu$ is the parameter regulating the range of the interaction
and $\rho$ is the equilibrium distance, and $r_{i,j}$ is the instantaneous distance for particles $i$ and $j$.

The other change we introduced is in the 4-body term. 
In the original version of HiRe-RNA, cooperative effects were calculated only for pairs of successive bases, as in the case of $i$ being bound to $j$ and 
$i+1$ to $j-1$. It can happen however that two base pairs are stacked on top of each other and therefore have a cooperative stabilization effect, even if they
are not successive along the chain.
This is typical of junctions where two helices can be perfectly stacked on top
of each other to the point of forming a seemingly continuous helix like in tRNAs and in many other large RNA molecules \cite{Holbrook2008}.
We modified the function to keep into account only the geometric placement of the two base pairs with respect to each other, and not their sequentiality.
Consider base $i$ bound to base $j$ and base $k$ bound to base $l$, with base $k$ stacking to base $i$ and base $l$ stacking to base $j$. 
The interaction energy is composed of four Gaussians:
\begin{equation}
E_{HB4} =-\varepsilon_{hb4}~ e^{-\gamma(r_{ij} - \rho_{ij})^2}e^{-\gamma(r_{kl} - 
\rho_{kl})^2} e^{-\delta(r_{ik} - r_{jl})^2} e^{-\lambda(r_{ik} + r_{jl} - 2r_0)^2}
\end{equation}
where $r_{i,j}$ is the instantaneous distance of particle $i$ and particle $j$, $\rho_{i,j}$ is the equilibrium distance for the specific hydrogen-bond 
(which varies depending on the base nature in the pair), $r_{i,k}$ and $r_{j,l}$ are the distances between particle $i$ and particle $k$ and particle $j$ and 
particle $l$ respectively, and $r_0$ is a cut-off distance for the interaction to vanish.
In essence this functions gives an additional energy gain if two bonds are formed simultaneously and if the bases are aligned in a ``square'' configuration.

As a minor change the 3 body function has been simplified from its original form
to gain stability and increase code performance:
\begin{equation}
 E_{HB3} = \varepsilon_{hb3}~ e^{-\eta (r_{i,j} - \rho_{i,j})^2} e^{-\eta(r_{i,k} - \rho_{i,k})^2}
\end{equation}
where particles $j$ and $k$ are both susceptible to pair with particle $i$. 

All these changes have been tested on the two benchmark molecules 1EOR and 1N8X used in the original publication.
Through MD simulations, the new version $\beta.2$ folds these molecules to their NMR structures within 2-3 \AA. 

\subsection*{Reconstruction procedure} \label{reconstruction}
The nucleobase being relatively rigid fragments, the full atomic configurations
of an RNA molecule can be well described from a small number of internal coordinates,
involving mostly the phosphate and sugar moieties\cite{Richardson2008}.
Most of this information is preserved in HiRe-RNA's representation.
To obtain a highly accurate reconstruction, we performed a statistical
analysis on the RNA structures present in the Nucleic Acid Database (NDB) \cite{Berman1992},
and determined which internal coordinates explicitly represented in our model
were most strongly correlate with the missing degrees of freedom.
Using this information, we are able to reconstruct high-quality
all-atom structures of RNAs from a corresponding coarse-grained representation.

\begin{figure}
  \begin{center}
  \includegraphics[width=0.8\textwidth]{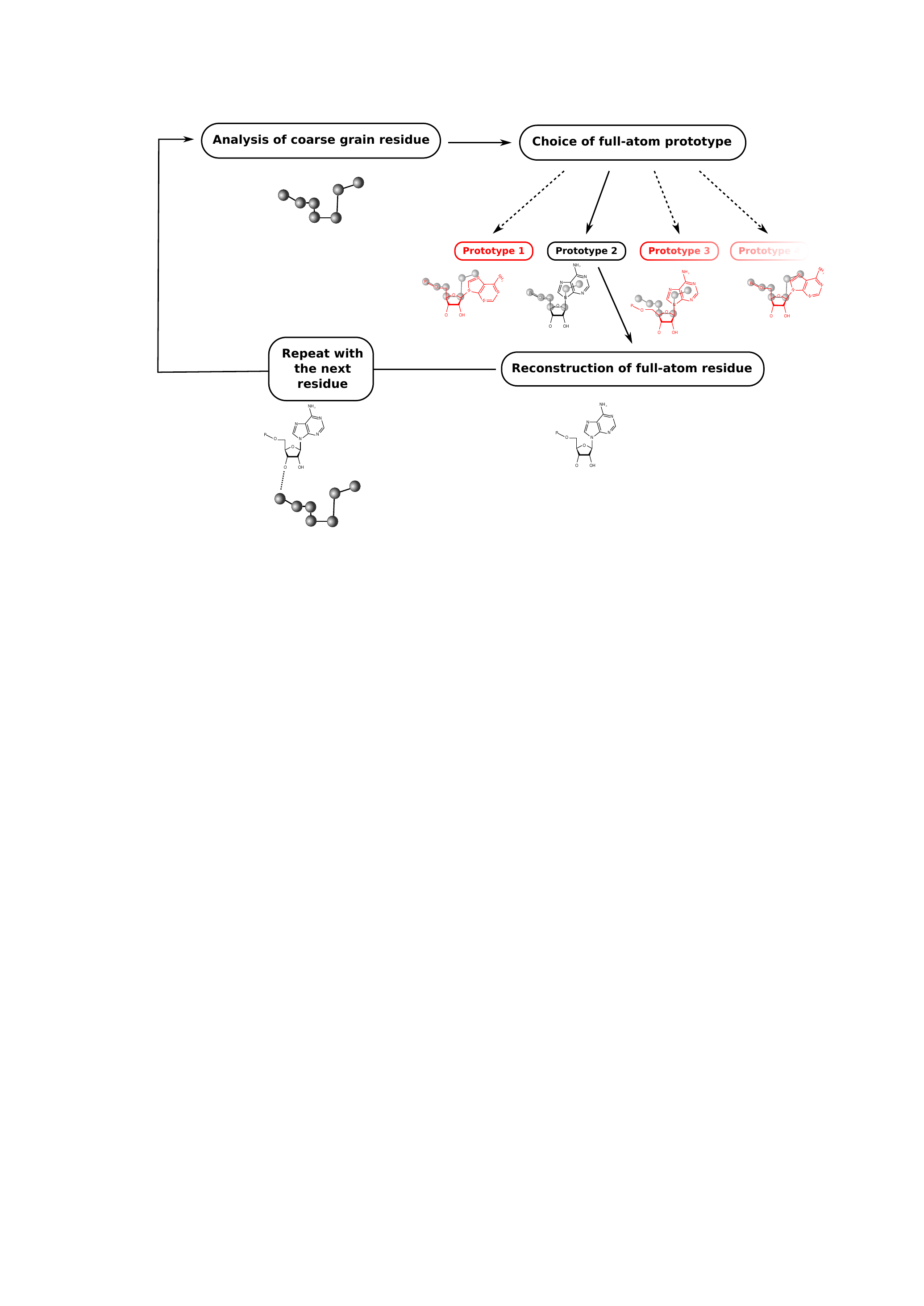}
  \caption{\label{rec_meth} Schematic of the reconstruction method used to convert from a coarse grain to an all-atom structure.}
  \end{center}
\end{figure}

Using the RNA conformers data presented in \cite{Richardson2008}, we selected a number of prototypes, represented in internal coordinates,
representing the most common motifs present in RNA (helix, loop and turn).
From coarse grain nucleobase bond lengths, angles and torsions, we can choose the best suitable atomistic prototypes,
and convert the internal coordinates into atoms positions \cite{Parsons2005a}.
This procedure is repeated for each residue (see Fig\ref{rec_meth}).
The overall scheme is particularly efficient given it uses neither a fragment search nor an optimisation procedure\cite{Heath2007,Rzepiela2010}.

To validate our approach, we generated coarse-grained models for 146 crystallographic
structures, and attempted to reconstruct back the fully atomistic structures.
The rmsd distribution between the original and reconstructed structures
has a mean of 0.18\AA , with a maximum value of 0.516 \AA, values that are well below the resolution of most crystallographic data on biomolecules.
This is a clear indication that the reduction scheme of HiRe-RNA well preserves the geometric information of an RNA conformation.

\section{Systems}\label{sys}
\begin{figure}
  \begin{center}
  \includegraphics[width=0.6\textwidth]{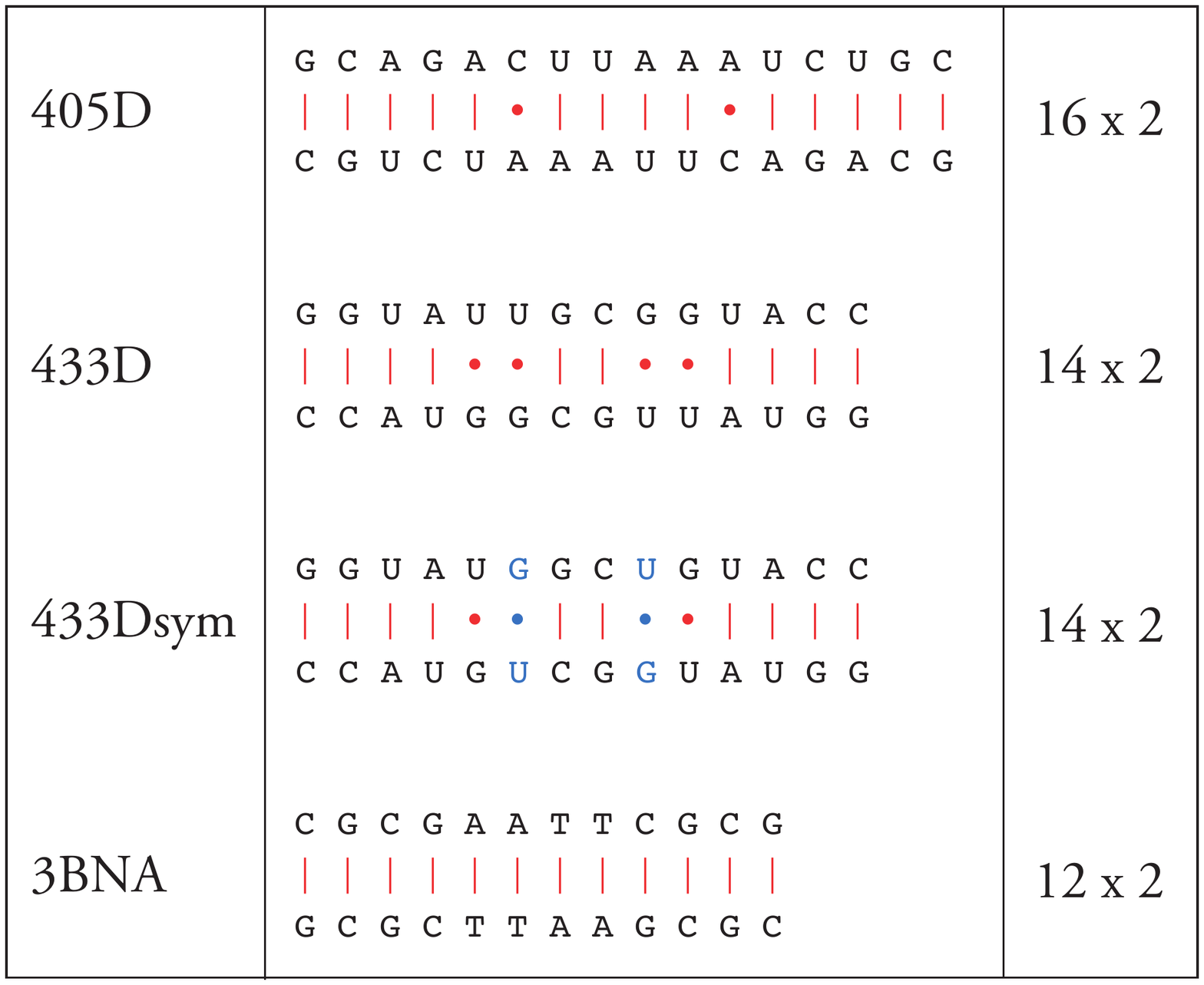}
  \caption{\label{fig_basepairs}Base pairings schemes for each one of the system analyzed. For 433Dsym in blue are the bases exchanged with respect to the wild form 433D.}
  \end{center}
\end{figure}

We investigate here molecules of size similar to the one presented in the original paper of HiRe-RNA, but with different topologies. 
We have chosen two RNA duplexes of a total of 32 and 36 nucleotides each that adopts a double helical structure and one DNA duplex of 24 nucleotides in total.
Both RNA double helices contain non-canonical base pairings, A$\cdot$C in one case, and G$\cdot$U in the other.
Figure \ref{fig_basepairs} shows the sequence and base pairing of all duplexes. 
We studied also a DNA double helix in the B-form, therefore a significantly different helix from the one formed by typical RNA, which 
adopts an A-form.

\subsection*{RNA double helices}
The two double helices are the ``Structure of a 16-mer RNA duplex with wobble C$\cdot$A mismatches'',
pdb code 405D \cite{Pan1998}, and  the ``Structure of a 14 bp RNA duplex with non-symmetrical tandem G$\cdot$U wobble base pairs'', 
pdb code 433D \cite{Trikha1999}.

\begin{figure}
  \begin{center}
  \includegraphics[width=1.1\textwidth]{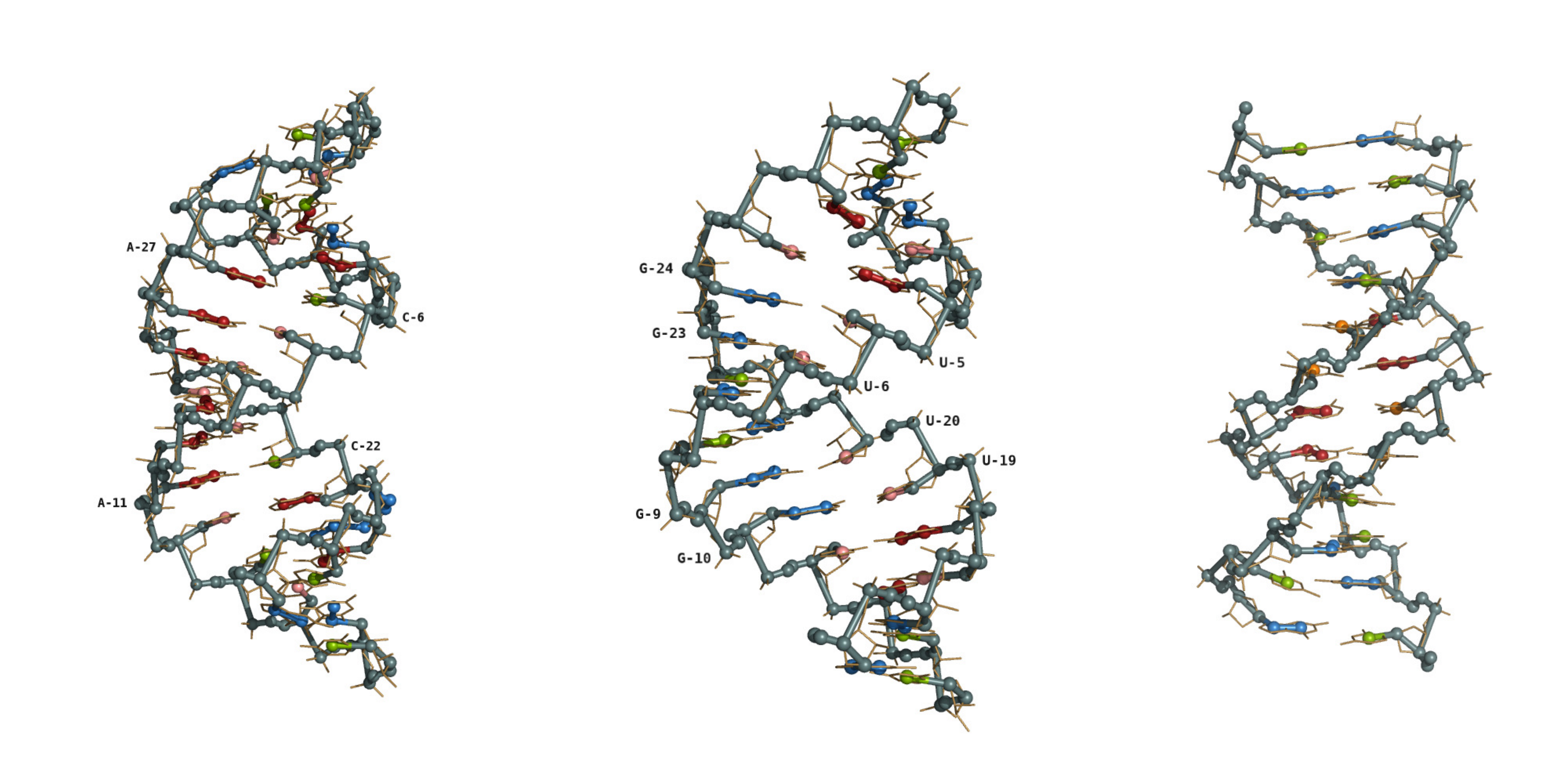}
  \caption{\label{fig_systems}The coarse-grained representation of three duplex systems superposed to the fully atomistic structures:
  left: 405D RNA, center: 433D RNA, right: 3BNA DNA. (Bases color coding: dark red:A, pink:U, orange:T, green:C, blue:G)}
  \end{center}
\end{figure}

The sequence of 405D is $\mathrm{r(GCAGACUUAAAUCUGC)_2}$ and its crystal structure has been 
resolved at 2.5$\mathrm\AA$ resolution \cite{Pan1998} (Fig. \ref{fig_systems} left).
It is characterised by two wobble A$\cdot$C pairs, between C6 and A27 and between A11 and C22 respectively. 
All other bases form canonical Watson-Crick (WC) pairs.
The presence of the mismatch alters the width of the grooves and provides a possible recognition site for proteins. 
A$\cdot$C mismatches are more rare than wobble G$\cdot$U pairings, but they are biologically important as they replace G$\cdot$U
pairs in ribosomal RNA and are found in some codon recognition of tRNAs. 

The sequence of 433D is $\mathrm{r(GGUAUUGCGGUACC)_2}$ and its crystal structure has been determined at a 2.1$\mathrm{\AA}$ 
resolution \cite{Trikha1999} (Fig. \ref{fig_systems} center). 
This duplex contains two successive non-symmetric wobble G$\cdot$U pairs, which deform significantly the helix with respect to a
regular A-helix. This deviation from the canonical RNA helix shape could play the role of a sequence-specific recognition site.
Experimental observations on the presence of G$\cdot$U motifs in ribosomal RNA, have shown that the non-symmetric tandem G$\cdot$U mismatch (5'-U-U-3'/3'-G-G-5')
is more rare than the symmetric motif 5'-U-G-3'/3'-G-U-5' in all kinds of secondary structure environments \cite{Gautheret1995}.
Is is also observed that the non-symmetric mismatch alters the stability of the 
molecule with respect to both an all WC helix, but also with respect to an helix containing the same mismatch but in 
the symmetric motif 5'-U-G-3'/3'-G-U-5'\cite{Trikha1999}.
To investigate the stability of the symmetric mismatch with respect to the non-symmetrical wild form, we created 
a new molecule, not existing experimentally, named 433Dsym, in which we have inverted one of the G$\cdot$U mismatches to obtain the symmetric motif.

\subsection*{DNA}
In our model there is no intrinsic distinction between ribose or deoxyribose given we model the sugar with only one bead.
The only changes we need to introduce in order to study DNA molecules are equilibrium values of the 
geometric parameters governing angles and torsions, which differ from RNA, and the relative values of the
energy of base pairings to keep into account that DNA couples almost exclusively through Watson-Crick
G$\cdot$C and A$\cdot$T.
We apply our model to a DNA double helix of size similar to the RNA duplexes.
The different hydrogen bond parameters are shown in Table 1.
Geometric parameters were taken from the standards values of B-DNA on the NDB.
We have chosen to study the 3BNA duplex\cite{Fratini1982} , which is a reference for B-DNA helix and whose size is of 12 nucleotides on each strand.\\
The sequence is $\mathrm{r(CGCGAATTCGCG)_2}$ (Fig. \ref{fig_systems} right).
Major and minor groves are clearly distinct, as typical for a B-helix, in contrast with an A-helix, where the two grooves 
are more similar in width. 

\begin{table}\label{table_HB_DNA}
\begin{center}
\begin{tabular}{ c c c c} 
base pair RNA & $\varepsilon_{hb2}$(kcal/mol) RNA & base pair DNA & $\varepsilon_{hb2}$(kcal/mol) DNA \\
\hline
\hline
A-U & 1.3 & A-T & 1.3 \\
A-G & 1.1 & A-G & 0.2 \\
A-C & 1.0 & A-C & 0.2 \\
G-C & 1.5 & G-C & 1.5 \\
G-U & 1.2 & G-T & 0.2 \\
C-U & 0.8 & C-T & 0.2 \\
\end{tabular}
\caption{HB 2-body coupling constants for RNA and DNA}
\end{center}
\end{table}

\section{Coarse-grained Simulation}\label{CGsim}
For all molecules we ran a set of Replica Exchange Molecular Dynamics (REMD) \cite{Chebaro2009} using a 
Langevin thermostat \cite{Spill2011}.
This is an enhanced sampling method allowing to sweep over several temperatures at once.
A preset number of Molecular Dynamic simulations (MD) are run simultaneously at
different temperatures, and at fixed time intervals a swap between configurations of neighbouring 
temperatures is attempted and accepted or rejected based on the Boltzmann probability calculated on the 
energy difference of the two configurations \cite{Derreumaux2007}.
In practice at the swapping time the energies of two replicas are compared and if the replica at the higher energy occurs
to have an internal energy lower than that of the lower temperature replica, the exchange between the two configurations 
is always accepted. In the opposite case, the exchange is accepted based on a Metropolis criterion. 
This procedure is carried out using a protocol recently presented in \cite{Chodera2007a} where a multitude of Monte Carlo step in attempted
replica swapping are performed at each preset time. 
This approach results in a great enhancement of the sampling while not altering the final statistical distributions of measurable
quantities.

Given that the force field of HiRe-RNA has not yet been
optimised, the absolute strengths of the interactions
do not reflect experimental values and only relative strengths are relevant.
Our temperature therefore looses for the moment its connection with the real
temperature and is expressed in terms of an arbitrary scale $\Theta$.

\subsection*{General Simulations Protocol}
For each duplex we generated REMD trajectories from a completely unfolded state. This
starting point was taken from a snapshot of a long simulation at high temperature starting from the
experimental structure. 
The configurations taken as the unfolded states all consist of two unpaired strands at an average distance comparable 
to the single strand length, and no inter-strand hydrogen bonds.

All simulation are carried out with spherical bounds of radius 65\AA, while the strand extension is between 70 and 60\AA~ for all molecules.
This relatively high concentration was chosen to minimise the encounter time of the two strands, which is basically a random walk in 3D,
and can be the most time consuming (and uninteresting) part of the simulation. 
 
The REMD consists of 24 replicas with temperatures in units of $\Theta$.\\
For the double helical RNA systems the temperatures are:\\
 220.0, 235.0, 246.677, 255.463, 264.563, 273.987, 283.746, 287.0, 290.0,
293.853, 297.0, 300.0, 302.0, 304.320, 308.0, 312.0, 315.160, 320.0, 326.386,
330.0, 338.011, 350.051, 362.433, 380.0. \\
For the DNA system the temperatures are:\\
215.00, 230.00, 250.000 262.818, 276.293, 290.459, 305.351, 321.006, 337.465, 345.00, 354.767, 362.00, 372.956, 377.00, 382.00, 387.00, 392.078, 397.00, 402.00, 412.180, 433.313, 455.530, 470.00, 490.885. \\
These values were chosen after several trial runs, in which we made sure to have an adequate exchange between replicas and a sufficient temperature
range to include the specific heat peak with a good margin on both sides.
Each replica was simulated between 100 and 600 ns, and the overall internal energy monitored. 
In the analysis we included only the portions of simulation where the internal energy had reached stability.
The convergence of the simulation is verified by the superposability of the thermodynamic properties computed separately on different 
portions of the simulation on one hand, and by the number of full excursions of the replicas on the entire temperature range, on the other. 
For the analysis, we monitored the number of native base pairs
and the overall rmsd from the experimental structure.
A base-pair is considered formed if the bonding two-body energy is at least half of its maximal value, 
i.e. $|E_{HB2}(\mathrm{bond})|\geq \frac{1}{2}|E_{HB2}(\mathrm{max})|$.
All conformations were clustered recursively as follows:
after computing the rmsd between all pairs of structures, 
we identify the structure with the largest number of neighbors using a 
cutoff of 2.5\AA; the center and all members of this cluster are removed and
the procedure is repeated until all states are clustered.
Finally, thermodynamic properties were extracted from REMD simulations
using the weighted histogram analysis WHAM \cite{Chodera2007a}.


For a molecule of 22 nucleotides, it takes 25 seconds per nucleotide per simulated nanosecond on an Intel 3.0GHz processor.
That is,  it takes 9 minutes to simulate 1ns, and 8 hours for 50ns.

\subsection*{Results}
For each REMD we have determined the specific heat curve, in which we can observe the thermodynamical transition
of assembly/dissociation of the duplex.
We have performed a cluster analysis of stuctures at each temperature to determine the most populated configurations.
Combining these two analysis allows us to draw the full picture of the molecule behavior at various temperatures.

In principle, if we only consider the energy of base pairing, we could expect a situation where low temperatures
are dominated by duplexes, since they maximise the number of pairings, and therefore minimise the internal energy, 
intermediate temperatures where the two strands are dissociated
but folded, and the melted state, where the strands are both dissociated and free-floating. 
However, for all molecules we find very neat assembly curves for the duplex with only one peak corresponding
to the assembly transition, while we never see hints of the presence of two separately folded single strands. 

\subsubsection*{405D}
\begin{figure}
  \begin{center}
  \includegraphics[width=1.0\textwidth]{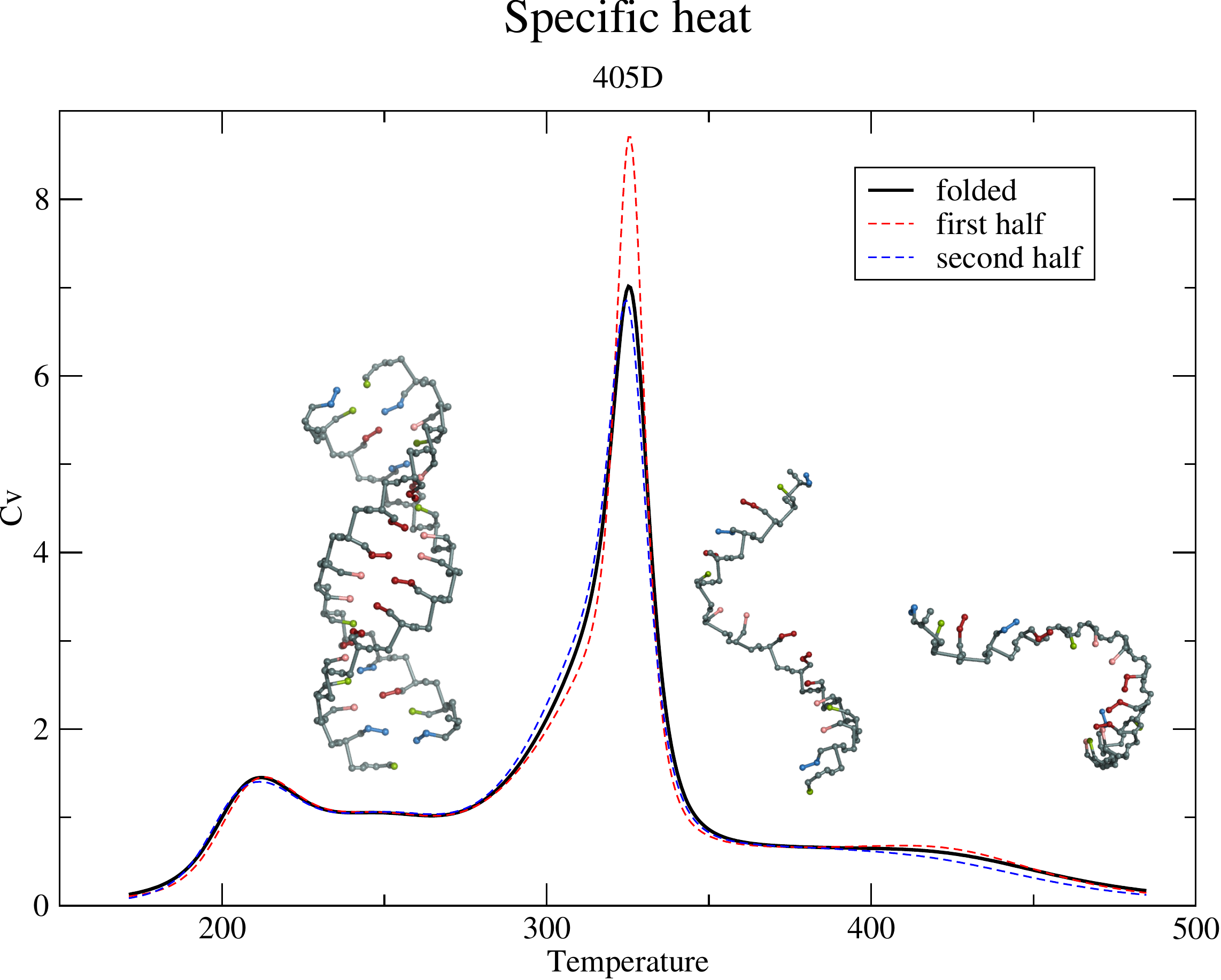}
  \caption{\label{fig_Cv_405D}Specific heat curve for 405D. The black curve is obtained for the whole trajectory having excluded the first 30 ns of the simulation. The two colored curves are obtained by considering the first half of the simulation (30-65ns: red) and second half (65-100ns: blue) respectively and by their superposition show the good convergence of the simulation. On the left of the peak the structure of the most populated structure at T3 (247 $\Theta$) and on the right a representative configuration at high temperatures, T22 (350 $\Theta$).}
  \end{center}
\end{figure}

Figure \ref{fig_Cv_405D} shows the specific heat curve for 405D obtained from a 100ns simulation.
The first 30ns were excluded from all analysis.
We can observe the good convergence of the simulation as verified by the superposability of the curve if we consider only half of the trajectory at the time.
To further assess the good convergence of the simulation we also monitored the migration of each replica over all temperatures, and we observed that each replica makes several dozens full temperature excursions in the simulated time.

The system exhibits a single melting transition at 325 $\Theta$, with no hints of the possible presence of two folded hairpins.
Below the melting temperature the native duplex structure is dominant, while above it the two strands are fully unfolded and behave independently.
In principle, because of the base complementarity, each single strand could fold back on itself to form an hairpin with a tetraloop,
each one still containing one of the two mismatches present in the duplex. 
The structures of the single strands for these duplexes are not known, but, knowing their sequences,
starting from a single strand random configuration we can obtain the structure of a folded hairpin.
We then ran REMD simulations from an unpaired state where the two single strands were folded 
upon themselves forming two stable hairpins.
This was done to test whether the double helix is indeed the most stable structure and to rule out that the duplex 
might be a very long lived metastable state. Indeed we find that in the temperature range of the duplex melting transition, and at our concentrations,
the folded hairpins open and reform the duplex.
We can then conclude that the duplex is the preferred configuration both energetically and entropically at the concentrations we are simulating,
 and that only by significantly increasing the dilution we could observe stable hairpins. 
This result is in qualitative agreement with the experimental evidence on kissing loops, where the configuration with the two hairping connected at the loop
can be a metastable state leading to an extended duplex \cite{Salim2012}.

\subsubsection*{433D, 433Dsym}
\begin{figure}
  \begin{center}
  \includegraphics[width=1.0\textwidth]{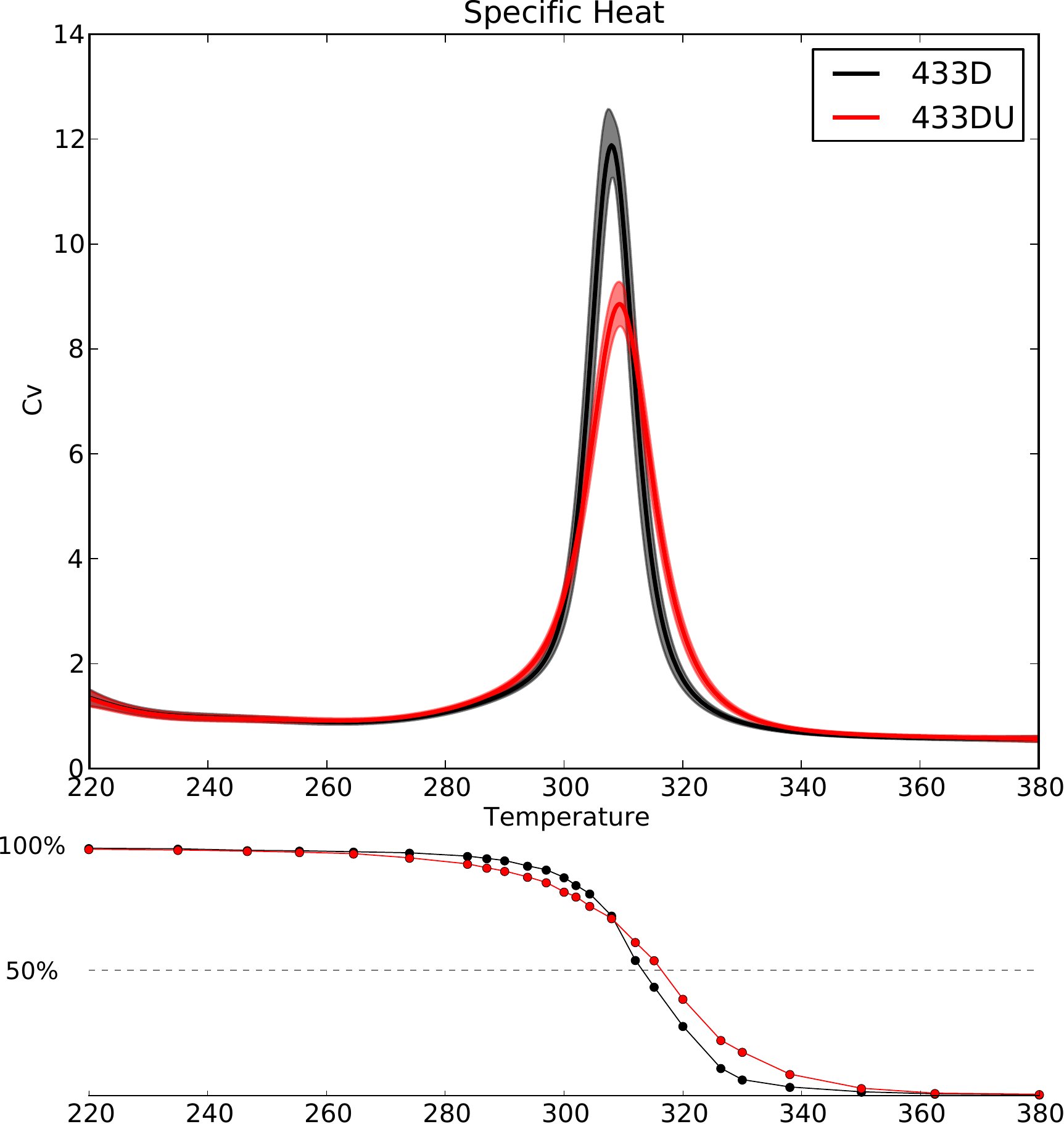}
  \caption{\label{fig_Cv_433D}Top: Specific heat curve for 433D (black) and 433Dsym (red). The shaded area around the middle line represents the unceirtanty of the measure calculated with the bootstrap method. Bottom: percentage of folded structures for the two systems as a function of simulation temperature.}
  \end{center}
\end{figure}

For these two systems our main question is whether or not we are able to capture the thermodynamic differences between the symmetric and the antisymmetric
tandem G-U mismatch. 
We expect these differences to be very subtle and we therefore need to cumulate a very large statistics. 
The two systems were simulated in 3 independent sets of REMDs of 600 ns each, and only the last 300ns of each simulation were used to compute statistics. 
Two of the simulations were started from unfolded configurations only, and one was started from a mixture of unfolded and folded initial configurations.
The specific heat curves are shown in figure \ref{fig_Cv_433D} for each system. 
For each curve we represent both the mean Cv curve as well as its error computed using a boot strap method to obtain a width of the curve representative 
of the uncertainty. 
For each temperature we also measure the overall percentage of folded structures for each system.
The two Cv peaks are superposable and if there are small temperature differences in their exact position, they are well within our uncertainty. 
Therefore, at this stage, we can not distinguish the two systems based on the position of the melting peak.
This result is somehow expected because our current model does not have a specific stacking term. 
We would expect stacking to contribute the most to the energetic difference as the vertical alignment with the preceding and successive bases would certainly be different when the two bases are inverted.
In our current model the energetic differences upon the base switch are due to the conformational changes induced on the backbone, and are therefore very subtle.

Although we can not distinguish the two systems based on the positions of the specific heat peaks, we observe significant differences in the width of the peaks,
with the peak for 433D more pronounced and narrower, than the peak of 433Dsym. 
This implies that 433Dsym has a transition over a wider temperature range than 433D. 
The difference in width at half maximum for the two peaks is of the order of 10 $\Theta$, which is above our uncertainty estimated at a few $\Theta$.
This observation is also supported by the data on percentage of folded structures.
Indeed 433Dsym starts having a significant proportion of folded structures at temperatures lower than for 433D and, symmetrically, it retains a fraction of folded structures at temperatures higher than for 433D. 
Our findings suggest a greater thermal stability of the sequence containing the symmetric mismatch. 
This result, although purely qualitative, is in agreement with the experimental evidence on tandem G$\cdot$U motifs stability. 
We can therefore conclude that despite our inability of detecting differences in the specific positions of the melting peaks, we can still capture significant differences in the molecules thermodynamical behavior. 

\subsubsection*{3DNA}
\begin{figure}
  \begin{center}
  \includegraphics[width=1.0\textwidth]{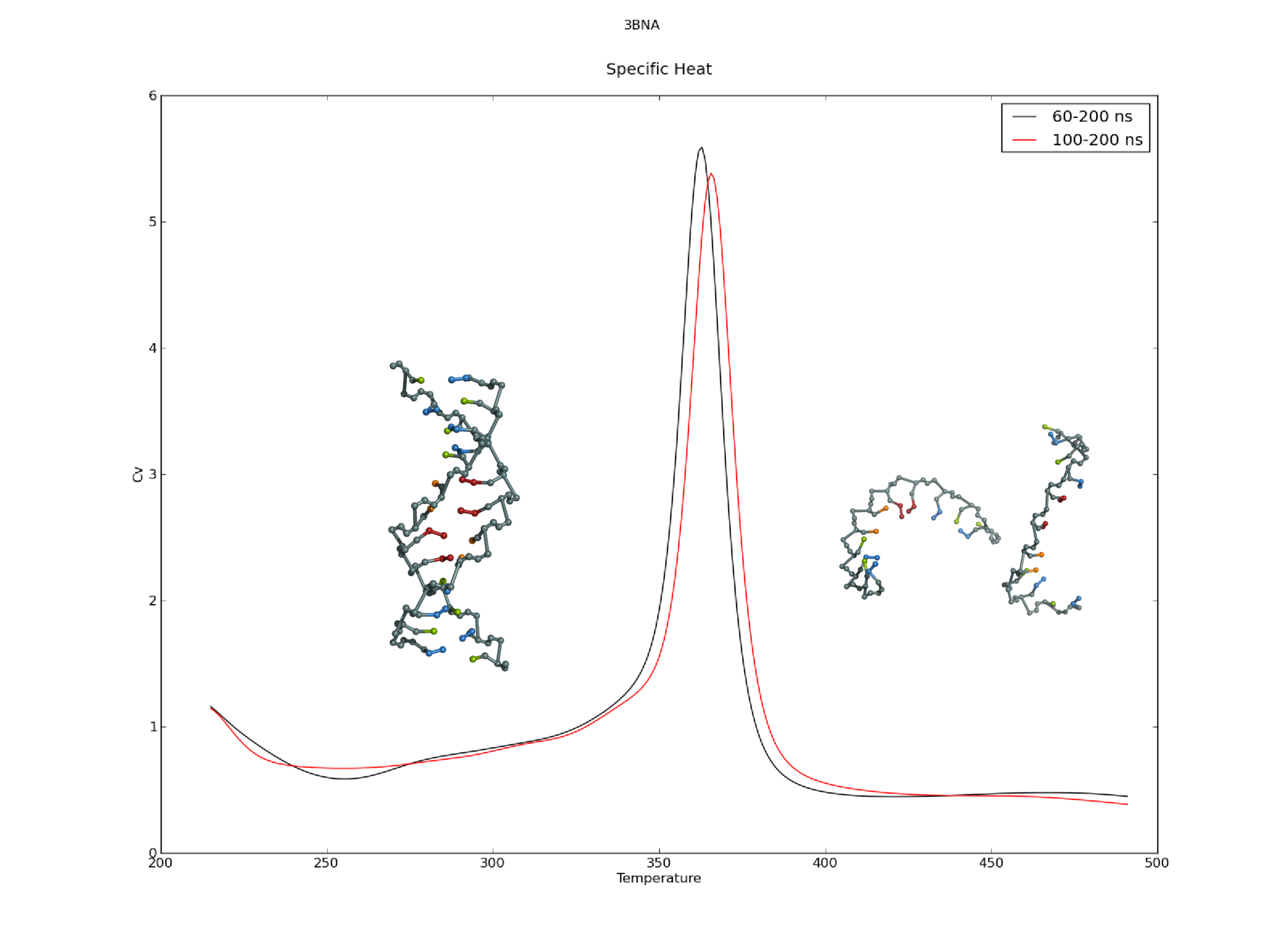}
  \caption{\label{fig_Cv_DNA}Specific heat curve for 3BNA.The representative structures at temperatures below the melting peak is the most populated cluster at T7 ( 305$\Theta$). It's n-rmsd is of 2.7\AA~ and all native base pairs are correctly formed. As we can visually observe, the major and minor grooves are clearly distinct as it should be for an A-form helix.}
  \end{center}
\end{figure}

For the DNA system the specific heat curve is shown in figure \ref{fig_Cv_DNA}. 
As for the RNA systems, we observe a very neat transition, characterized by one single peak corresponding to a melting transition.
We can observe that the DNA molecule melts at higher temperatures than the RNA duplexes. 
This is expected from the base content of the DNA where all base pairs are canonical WC couplings. \\
Cluster analysis shows the presence of a single dominant cluster at low temperatures with the correct native base pairing, and with an rmsd of 2 to 3 \AA~ 
from the experimental structure. The cluster gradually drops in population approaching the melting peak.

\section{Atomistic simulations}\label{FAsim}
The main goal of carrying on atomistic simulations on our systems is to compare the behavior of simulations starting from experimental structures with simulations starting from structures reconstructed from coarse-grained predictions.
If the two sets of simulations are compare well, we are then justified to use the coarse-grained model to perform folding and assembly simulations of molecules for which the native state is unknown, and then refine the results via atomistic simulations. 

For the three double strands systems for which the experimental structure exists (the RNAs 405D and 433D, and the DNA 3BNA), 
the reconstruction is carried out following the procedure outlined in section \ref{reconstruction} from a low temperature equilibrium configuration of 
the REMD simulations.
To verify if the simulations from the experimental structure and the corresponding reconstructed structure are comparable in general behavior, we monitor 
the time evolution of the rmsd with respect to the experimental structure, computed on all heavy atoms of all internal nucleotides (i.e. excluding from calculations both ends of the chains), the time evolution of the number of inter-base hydrogen bonds, and we compare the distributions of the most populated clusters. 
Hydrogen bonds are calculated with the program VMD \cite{Humphrey1996} with the criteria of a donor-acceptor distance of less than 3\AA~ and of an angle of less than 20 degrees.
Since the instantaneous number of hydrogen bonds is subject to large fluctuations, due to the sharp definition used in calculating the presence
of a bond, we compute instead a running average over a 100 frames and use this quantity for the comparisons between the two simulations.
Clusters are computed using the gromacs \cite{Hess2008} program g\_cluster with a cutoff of 2 \AA.

\subsection*{All-atom Reconstructed Molecules}
First we compare the starting structures for the simulations. 
For each system we have 4 possible structures: the experimental crystal structure (C), the experimental structure relaxed after thermalization (CX), the reconstructed structure (R), and the reconstructed structure after thermalization (RX). 
When we consider the reconstructed structure R in comparison with C, the rmds are low (2.75\AA\ for 3BNA, 2.55\AA\ for 433D and 3.7\AA\ for 405D), but 
we notice a significant discrepancy in the number of inter-base hydrogen bonds.
This suggests that some of the bases of the reconstructed structures are not well aligned despite the general shape of the molecule being very similar.
We also notice that none of the crystal structures has a number of inter-base hydrogen bonds close to the value one would expect in theory adding the number of bonds of each kind of base pairs (2 for AU, GU, AC when considering a WC/WC binding, and 3 for GC). The number of inter-base hydrogen bonds is around 50\% for all systems instead.

When we look at the relaxed structures all major differences disappear. 
Rmsds for all three system are lower (2.60\AA\ for 3BNA, 1.82\AA\ for 433D and 2.8\AA\ for 405D), but the number of hydrogen bonds differs of at most 5 bonds (433D), which accounts for 15\% of all possible bonds.
It is then clear that comparisons are meaningful only on relaxed structures for both the experimental and reconstructed configurations.

\subsection*{All-atom Simulations Protocol}
For all-atom simulations, we used the parmbsc0 forcefield \cite{Perez2007a}
as implemented in amber 11. We used the TIP3P model for water \cite{Jorgensen1983a},
and the long-range electrostatic interaction were treated with the particle mesh Ewald method,
using a 10\AA\ real space cutoff.
SHAKE was used to constrain all bonds involving hydrogen atoms, and a 2 fs timestep was used.
Temperature was kept at 300K using a Langevin thermostat and a 5 ps$^{-1}$ collision frequency.
Pressure coupling used the Berendsen algorithm with isotropic position scaling,
and a relaxation time of 2 ps. The target pressure was kept at 1 bar.

All simulations were done using the same protocol :
\begin{itemize}
  \item the system's charge was neutralized by adding Na$^{+}$ ions,
  \item a truncated octahedron box was used, with a minimal buffer distance of 10 \AA\
  between the box's boundary and the RNA molecule. Water molecules using the TIP3P model were then added to the box.
  \item The system was then subjected to energy minimization,
  \item a short MD run 0.2ns was used, with restraints on the solute, in order to equilibrate the solvent,
    and the soft restraints used on the solute allowed reconstructed structures to relax to more stable states,
    i.e better stacking of neighboring bases, and improved orientation in a base pair.
  \item With all restraints removed, we did a thermal equilibration, followed by pressure equilibration for a total of 0.8ns.
  \item Finally, the production simulations were run for 10 ns in the NPT ensemble.
\end{itemize}

\subsection*{Experimental vs. reconstructed structure MD}
For all three systems, we are now going to compare atomistic MD simulations ran starting from the experimental structures (E-sim) and 
simulations ran from the reconstructed structure (R-sim) obtained from the most populated low temperature cluster of the coarse-grained simulation, 
for the two RNA double helices 405D and 433D, and for the DNA 3BNA.

\begin{figure}
  \begin{center}
  \includegraphics[width=1.0\textwidth]{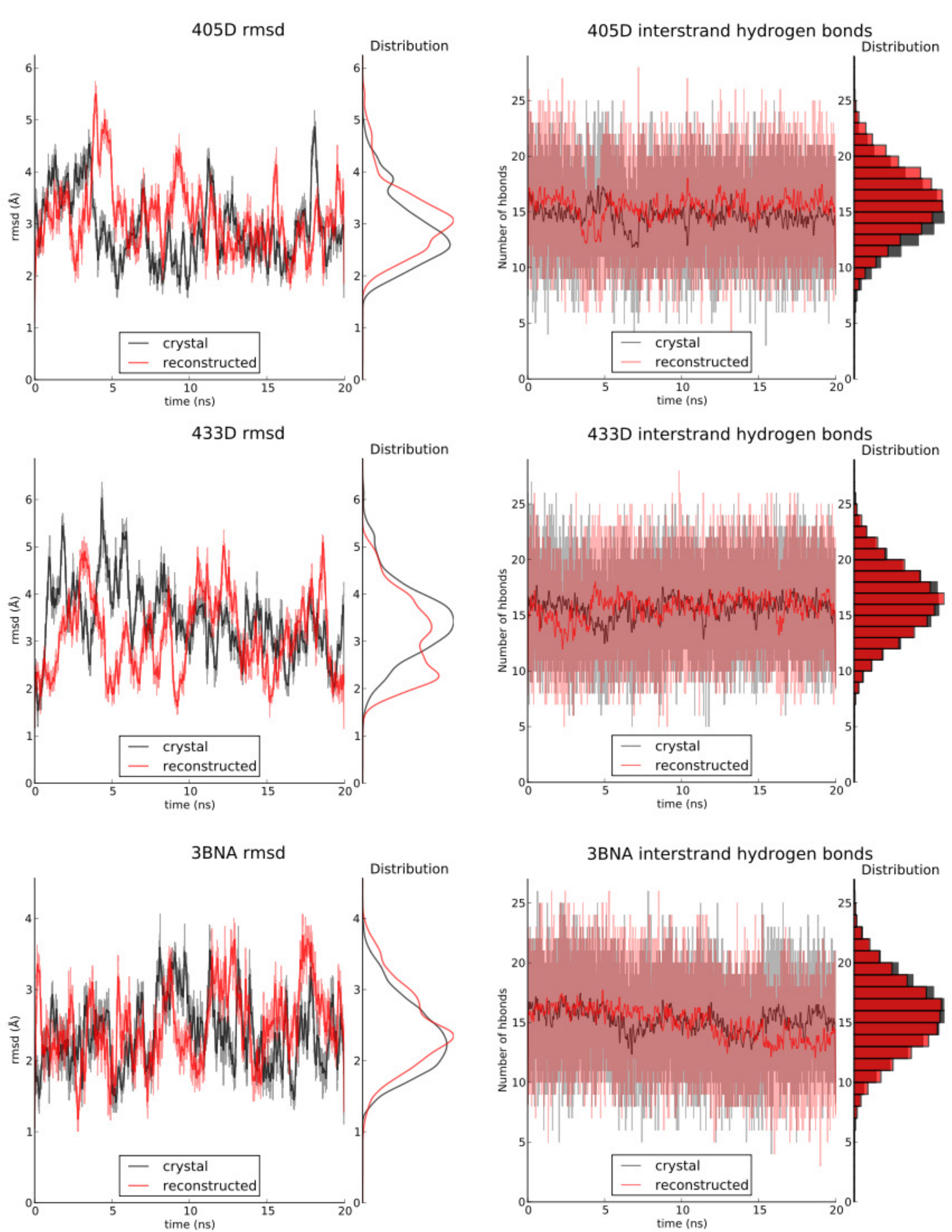}
  \caption{\label{FA_sim} Distributions of n-rmsd and number of instantaneous inter-strand hydrogen bonds for each atomistic systems in the two set of simulations, from the experimental structure and the reconstructed structure.}
  \end{center}
\end{figure}

All three sets of simulations exhibit a very similar behavior when starting from the experimental structure and from the reconstructed structure.
We observe no significant difference in neither the time evolution of the rmsd nor in the number of inter-base hydrogen bonds. 
This can be observed also from a comparison of the respective distributions of these two quantities over the whole simulation, which are superposable in both instances (Fig. \ref{FA_sim}).
In both E-sim and R-sim, the number of window-averaged inter-base hydrogen bonds (IBHB) fluctuates of 3 or 4 bonds around the mean value (i.e. a fluctuation of about 20\% with respect to the average number of bonds present during the simulation, or of 10\% of the overall possible number of bonds), while the instantaneous values vary much more significantly, with fluctuations of the same order as the mean value.
This is also in part due to the sharp cutoff used to evaluate the presence of a hydrogen bond.
For example, the crystal structure of 405D is characterized by 21 hydrogen bonds between the bases, but throughout the simulation the number of instantaneous IBHB fluctuates symmetrically around  the mean value of 16, spanning from as little as 6 to a maximum of 27.
A very similar result is obtained from the reconstructed structure with a mean value of IBHB 16 as well, a lowest value of 3 and a highest value of 25. 
Similar results are found for the other two systems as shown on the right hand side of figure \ref{FA_sim}.
However, the overall shape of the molecule does not change significantly, as shown by values of rmsd with respect to the experimental structure that rarely goes above 5\AA (left hand side of Fig. \ref{FA_sim}).
In the time of our simulations, what we observe are possible local rearrangements of the bases, that can slightly bend and/or rotate about their plane and about the backbone. 

\begin{figure}
  \begin{center}
  \includegraphics[width=1.0\textwidth]{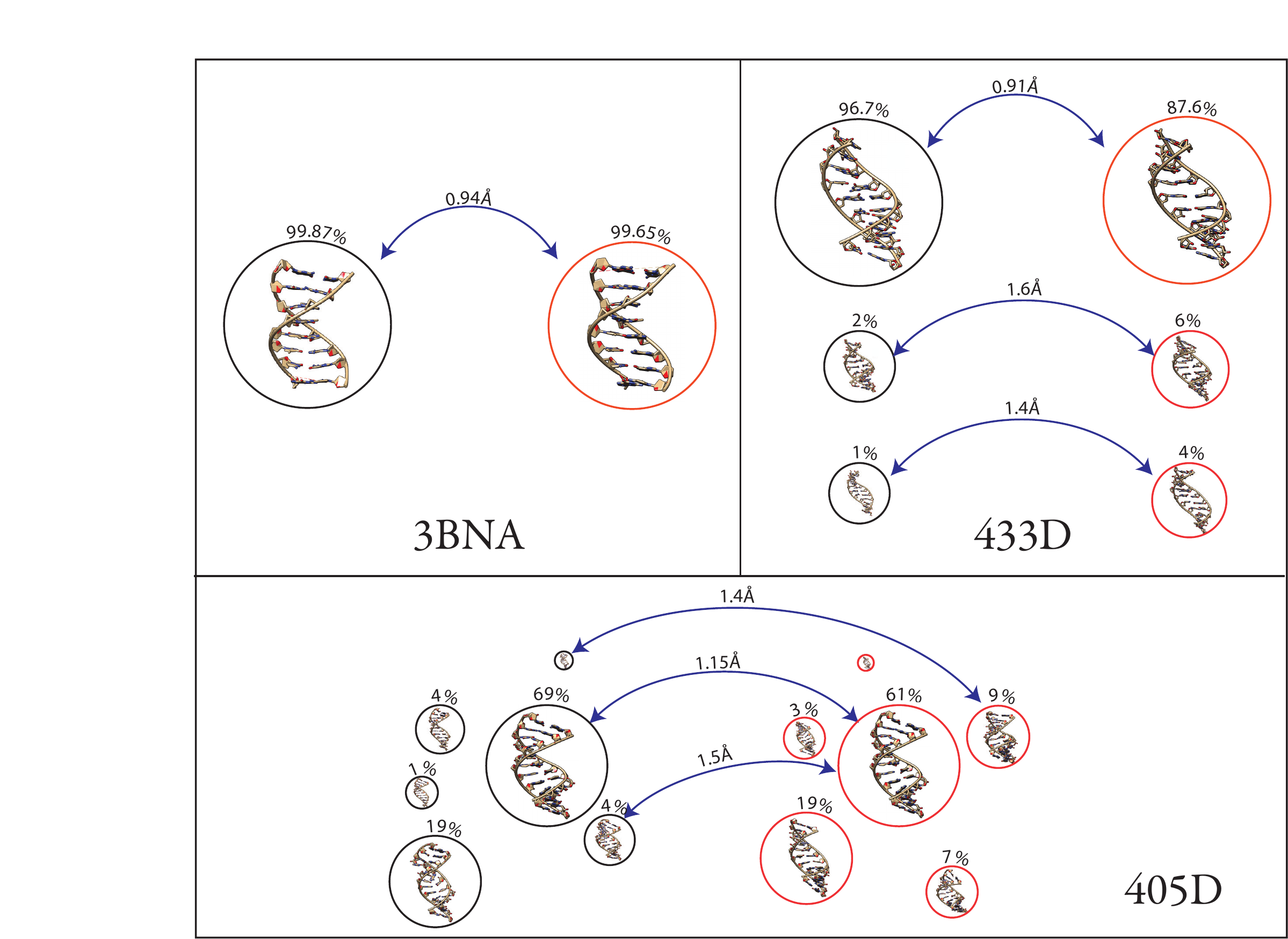}
  \caption{\label{FA_clusters} Cluster comparisons between the E-sim (black) and the R-sim (red) for the three systems. 
The percentage on top of each cluster indicates relative cluster population and the arrow between a black and a red cluster indicates the rmsd between corresponding cluters in the two types of simulations, measured from the cluster centers.
}
  \end{center}
\end{figure}

For both E-sim and R-sim we notice that the RNA duplexes show more variability than the DNA duplex. 
Clustering analysis shows only one significant cluster for the DNA system (native: 99.8\%, reconstructed: 99.6\%), while for the RNA systems, one cluster is clearly dominant, but other smaller clusters are also present (Fig. \ref{FA_clusters}).
The most populated cluster of 433D is at 96.6\% for the native simulation and at 87.6\% for the reconstructed simulation, and of 405D at 69.2\% for the native simulation and at 61.3\% for the reconstructed.
When we compare the clusters for both sets of simulations, we see that in all instances the most populated cluster found in the native state simulations corresponds to the most populated cluster of the reconstructed simulation, as it is shown by crossed rmsds of about 1\AA, while the clustering is performed using a 2\AA~ cutoff. 
For 3BNA only one cluster is present in both simulations and the centers of the clusters are 0.94\AA~ apart.
For 433D the center of the most populated clusters on both simulations are 0.91\AA~ apart.
The other, less populated, clusters have a clear correspondence between the two simulations, with the second most populated cluster of one simulation exhibiting the lowest rmsd with the second most populated cluster of the other simulation (1.6\AA~), and the same happening also for the third most populated clusters (1.4\AA).
For 405D the most populated clusters are 1.15\AA\ apart, but we cannot find a clear cut correspondence between other clusters.
This is the system with the most variability, as reflected through both sets of simulations. 

\subsubsection*{Grooves analysis}

Using Curves++ and Canal\cite{Lavery2009}, we performed an analysis of the groove widths for atomistic simulations using the crystal and the reconstructed structures.
The minor groove widths, although varying for the different systems, does not exhibit large variations, and is nearly identical
between simulations from the crystal and reconstructed structures.

\begin{figure}
  \begin{center}
  \includegraphics[width=0.6\textwidth]{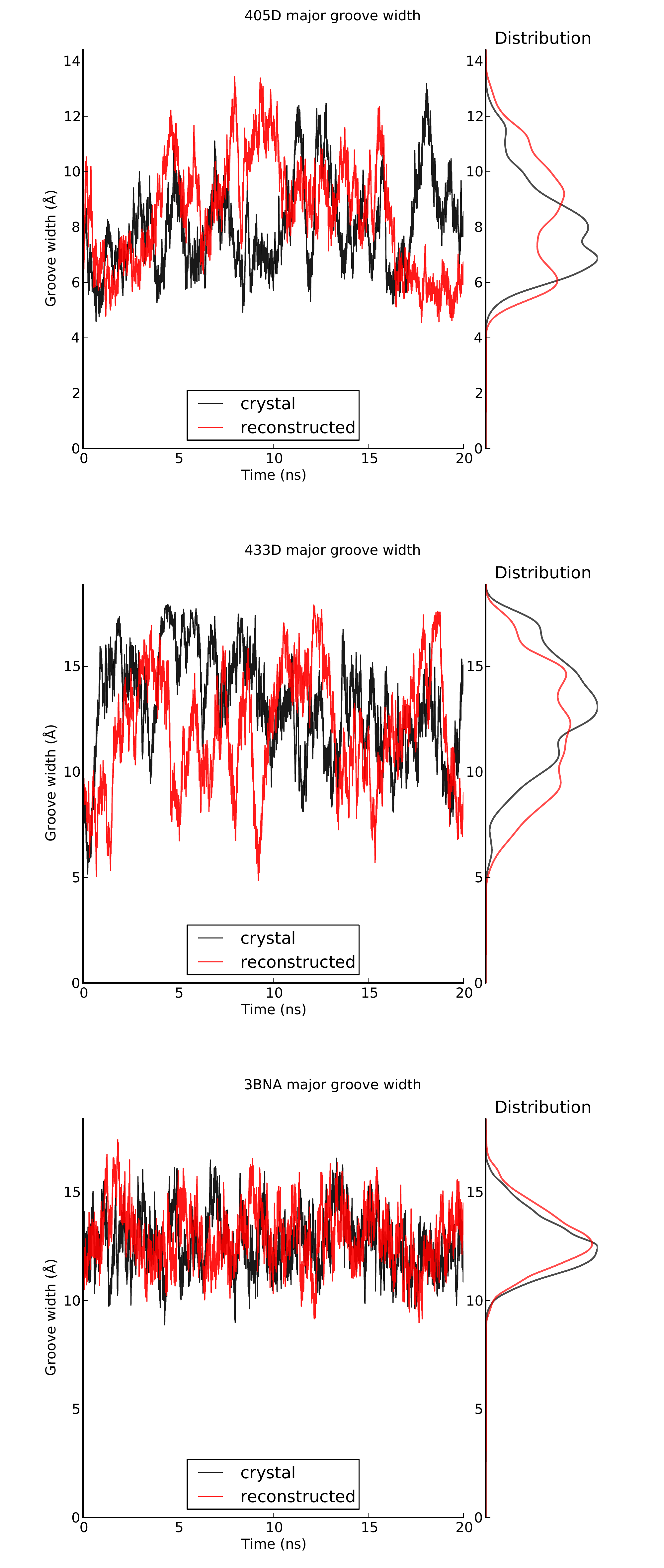}
  \caption{\label{cmb_maj_groove} Time series evolution and distribution of the major grooves.}
  \end{center}
\end{figure}

The major groove widths, presented in figure~\ref{cmb_maj_groove} shows much larger variations, although the values stays similar between crystal and reconstructed structure.
For the two RNA systems, the values are well in range of the usual values for RNAs
(one should note that the values for groove width calculated by Curves++ are decreased by 5.8 \AA , to take into account the Van der Waals
surface from the backbone).
For 433D, the major groove width from the starting simulation is among the lower values observed during the simulation.
The values sampled during our simulations falls in the range of values present in NMR structures,
with X-ray structures generally having narrower major grooves, as presented in ref~\cite{Tolbert2010}.

\section{Conclusions}
The study of large conformational changes in nucleic acids, such as folding or assembly, remains an open challenge in the computational field because
of the large size of the molecules and of the long times scales on which such transitions occur.
At the time being, the only viable path to approach sizable nucleic acid systems is a drastic reduction of the degrees of freedom that are considered.
The choice of what to retain in the model largely depends on the biological question one wants to address.
This reflects in the existence of several coarse-grained models for RNA and DNA, each one with its own peculiarities both in the type of representation
adopted, and in the force fields implemented. 

In this work we have presented a new application of our high resolution, fully flexible model for RNA and DNA.
A specific feature of our model is the high resolution of the nucleotides, that we represent with 6 or 7 beads, instead of 2 or 3 like in most other models. 
This degree of accuracy in the geometrical description allows us to have a system in which we do not have to impose constraints to hold 
the molecule in native-like conformations, and makes our model suitable for studies of folding and assembly.
All configurations can be explored, even if they depart significantly from the regular double helix and the canonical Watson-Creek pairing. 
Yet, the reduction, is still very significant and allows us to perform simulations in which we can indeed observe large conformational changes.
As we have shown here as a proof of principles, with our model we can investigate both dynamical and thermodynamical aspects of double helical assemblies.
We can compare the thermodynamics of molecules with the same base composition, but with different sequences, and investigate the effect of sequences on the 
melting curves of the duplexes. 
This result opens to the possibility of more detailed studies in which the sequence dependence of the thermodynamical and dynamical behavior can be studied systematically.

No matter the model, when adopting a coarse-grained approach, there is always a level of detail that is lost. 
We have developed a reconstruction algorithm that allows us to convert coarse-grained structures back to full atomistic details. 
This reconversion is easily feasible in our model thanks to the high level of resolution we adopt and because the nucleotides are relatively rigid objects:
such reconstruction would not have been so successful had we adopted a more reduced model of 2 or 3 beads only.
We have shown here that atomistic simulations carried out starting from experimental structures and from structures reconstructed from coarse-grained molecules are substantially equivalent.
This validates the possibility, for molecules for which an experimental structure is not known, to use a coarse-grained representation and to run (RE)MD simulations from an arbitrary configuration to obtain a low temperature stable structure, and then, after atomistic reconstuction, use the power of atomistic simulations to investigate all details, once the heavy duty job of folding or assembly has been carried out by the coarse-grained model.

Our coarse-grained model still needs to be optimized in its parameters and functional form (to include explicit stacking potentials for example) therefore the results we present here are only to illustrate its potential and versatility should not to be considered accurate in their quantitative details. 
In particular a correspondence between our temperature scale and real temperatures still needs to be drawn. 
For this purpose we now have an ongoing collaboration with experimentalists to provide us with data directly comparable to our simulation results. 
As a final remark, it should be noticed that the need for further refinement is also true of DNA and RNA atomistic models, where the force-fields are in continuous development \cite{Lagant1991,Yildirim2012,Hart2012,Krepl2012}.

\bibliographystyle{unsrt}
\bibliography{publis}

\newpage

\end{document}